\documentstyle[12pt]{article}
\newcommand{\be}{\begin{equation}}
\newcommand{\ee}{\end{equation}}
\newcommand{\ba}{\begin{eqnarray}}
\newcommand{\ea}{\end{eqnarray}}
\newcommand{\baa}{\begin{eqnarray*}}
\newcommand{\eaa}{\end{eqnarray*}}
\newcommand{\lab}[1]{\label{#1}}

\newcommand{\dis}{\displaystyle}

\newcommand{\bhat}{\hat{\beta}}

\begin{document}
{\pagestyle{empty}
~~\\

\vskip 2.0cm
{\renewcommand{\thefootnote}{\fnsymbol{footnote}}
\centerline{\large \bf Numerical Comparisons of Three Recently Proposed 
            Algorithms} 

\vskip 0.5cm
\centerline{\large \bf in the Protein Folding Problem} 
}
\vskip 2.0cm
 
\centerline{Ulrich H.E.~Hansmann$^{\#,\dagger,}$
\footnote{\ \ e-mail: hansmann@ims.ac.jp} and Yuko Okamoto$^{\dagger,}$
\footnote{\ \ e-mail: okamotoy@ims.ac.jp}} 
\vskip 1.5cm
\centerline{$^\#${\it Swiss Center for Scientific Computing (SCSC)}} 
\centerline{{\it  Eidgen{\"o}ssische Technische Hochschule (ETH)
 Z{\"u}rich, 8092 Z{\"u}rich, Switzerland}}
\vskip 0.5cm
\centerline {$^{\dagger}$ {\it Department of Theoretical Studies, 
Institute for Molecular Science}} 
\centerline {{\it Okazaki, Aichi 444, Japan}}

\medbreak
\vskip 3.5cm
 
\centerline{\bf ABSTRACT}
\vskip 0.3cm
   
We compare numerically
the effectiveness of
three  recently proposed algorithms, multicanonical simulations, 
simulations in a $1/k$-ensemble, and 
simulated
tempering, 
for the protein folding
problem. For this we perform simulations with high statistics for one of
the simplest peptides, Met-enkephalin. While the performances of all three 
approaches is much better than traditional methods,
we find that the differences among the three are only marginal.

\vfill
\newpage}
 \baselineskip=0.8cm
\noindent
{\bf INTRODUCTION} \\
While  there has been a considerable progress in numerical simulations
 of biological
macromolecules  over the last thirty years (for a recent review see, 
for instance, Ref.~\cite{VNS}), the
prediction of their low-energy conformations from the first
principles remains a formidable task. 
The numerical calculation of physical quantities
depend on how comprehensively the phase space of the system is explored.
However,  the
energy landscape of the system of peptides and proteins 
is characterized by a multitude
of local minima separated by high energy barriers. At low temperatures
traditional  Monte Carlo and molecular dynamics simulations tend to get trapped
in one of these local minima. Hence, only small parts
of the phase space are sampled (in a finite number of simulation steps) and
physical quantities cannot be calculated accurately. A common approach 
to alleviate this multiple-minima problem is to identify the global minimum
in the {\it free} energy (which should corrospond to the native
conformation of a protein) with the lowest potential energy conformation,
ignoring entropic contributions, 
and to search for this conformation with powerful optimization techniques
such as 
 Monte Carlo with minimization, \cite{LS} 
 genetic algorithms, \cite{Fo} 
or simulated annealing.\cite{SA}  
Since simulated annealing was first introduced to the protein 
folding problem,\cite{SA1}-\cite{SA4} the effectiveness was tested 
with many peptides and proteins, and it is now one of the
most popular optimization methods used in the field (for a recent review
see Ref.~\cite{SA5}).

It is claimed that a variety of recently proposed methods like the  
{\it multicanonical approach},\cite{MU,MU3}  {\it entropic
sampling},\cite{ES} {\it simulated tempering},\cite{L,MP} 
and simulations in a so-called 
$1/k$-ensemble\cite{HS} allow a much better sampling of 
the phase space than previous methods. 
Actually, two of these methods, 
{\it multicanonical approach} and
{\it entropic sampling}, are mathematically identical as shown
in Ref.~\cite{COMM}. 
The {\it multicanonical algorithm} was 
first applied to the protein folding problem in Ref.~\cite{HO}.
In Ref.~\cite{HO94_3} the authors showed   
that this new ansatz yields 
 indeed to an improvement over simulated annealing.
The method was also applied to the studies of the coil-gobular
transitions of a model protein \cite{HSp} and the helix-coil transitions of
homo-oligomers of nonpolar amino acids \cite{HO95a}.  Simulated
tempering was applied to the study of a model heteropolymer.
\cite{IP} 

Applications of 
simulated tempering and the $1/k-ensemble$
 to the protein folding problems are still missing. 
Also missing is a
quantitative comparison of the effectiveness of these three new methods. In
the present article we try to fill in this gap. By simulating one of the
simplest peptides, Met-enkephalin, with high statistics, we study the numerical
performances of the three methods.
In particular, the efficiency of finding the global-minimum-energy
conformation and calculating thermodynamic quantities is studied in
detail.

\noindent
{\bf METHODS}\\
{\bf Algorithms}\\
Although the algorithms are explained in the original papers, we
briefly summarize here the ideas and implementations of the three
approaches for completeness.

Simulations in the canonical ensemble weight each configuration with
the Boltzmann factor
$w_{B}(T,E) =  e^{-\hat{\beta} E}$
and yield the usual 
bell-shaped probability distribution of energy at temperature
$T$: 
\begin{equation}
P_B(T,E)~ \propto ~n(E)~w_B(T,E)~=~n(E)~e^{-\hat{\beta} E}~,
\end{equation}
where $n(E)$ is the density of states (spectral density) and
$\hat{\beta} = 1/k_B T$
is the inverse temperature (we set the Boltzmann constant $k_B$
to unity hereafter).  

Using local updates, the probability for the system to cross 
an energy barrier is
proportional to $e^{-\hat{\beta}\Delta E}$ in the canonical ensemble.
Here, $\Delta E$ is the height of the energy barrier. Hence, at low
temperatures (large $\hat{\beta}$) simulations will easily 
get trapped in one of the
local minima and very long simulations are necessary to
calculate thermodynamic quantities accurately. In general, one can think of 
two strategies
to alleviate this problem. 
First, one can look for improved updating scheme.  For instance,
a kind of {\it global} updating scheme called the
cluster update method \cite{CA,CA2} is popular in spin systems.
No such global updates are known for proteins.
Secondly, one  can perform the simulation in a 
"{\it generalized ensemble}", where the above difficulty does not arise.
The latter  approach is the one that is investigated in the present article.
 
In the {\it multicanonical approach} \cite{MU,MU3} configurations with
energy $E$ are updated with a weight:
\begin{equation}
 w_{mu} (E)\propto \frac{1}{n(E)} = e^{-S(E)}~,
\lab{eq2}
\end{equation}
where 
\be
S(E) = \log n(E)
\lab{eq3}
\ee
is the microcanonical entropy.  A simulation with this weight factor
will produce a 
uniform distribution of energy,
\begin{equation}
 P_{mu}(E) \propto n(E)~w_{mu}(E) = {\rm const}~,
\end{equation}
resulting in
a free random walk in the energy space.  This allows the simulation to escape
from any energy barrier, and even regions
with small $n(E)$ can be
explored in detail. 

Unlike in a canonical simulation the weight $w_{mu} (E)$ is not 
{\it a priori} known (in fact, the knowledge of the exact weight is equivalent
to obtaining the density of states $n(E)$, i.e., solving the system), and 
one needs its estimator for a numerical 
simulation.  Hence, the multicanonical ansatz consists of
three steps:
In the first step the estimator of the multicanonical weight factor
$w_{mu} (E)$ is calculated.
Then one performs  with this weight a multicanonical
simulation with high statistics. In this way information  is collected over 
the whole energy range.
We remark that there is no explicit temperature dependence in simulations
of the multicanonical ensemble. However,  from this simulation one can not
 only locate the energy global minimum but also obtain the
 canonical distribution at any inverse 
temperature $\hat{\beta}=\frac{1}{T}$
 for a wide
 range of temperatures by the re-weighting techniques:\cite{FS}
 \begin{equation}
 P_B(T,E) \propto P_{mu} (E)~w^{-1}_{mu} (E)~e^{-\hat{\beta} E}~.
\label{erw}
\end{equation}
This allows one to calculate the expectation value of any physical 
quantity ${\cal O}$ at temperature $T$ by
\begin{equation}
< {\cal O} >_T ~= \frac{\displaystyle{\int dE~ {\cal O} (E) P_B(T,E)}}
                       {\displaystyle{\int dE~ P_B(T,E)}}~.
\lab{rewt}
\end{equation}
 
The crucial step is the calculation of the estimator
for the multicanonical weight factor $w_{mu} (E)$. It is obtained
by an iterative procedure. 
The improved estimator of the multicanonical
 weight for the $i$-th iteration is 
 calculated from the histogram of energy distribution 
$P^{(i-1)}_{mu} (E)$ and the weight $w_{mu}^{(i-1)}$ of the preceeding
 simulation as follows:
\begin{equation}
w^{(i)}_{mu} (E) = \frac{w^{(i-1)}_{mu} (E)}{P^{(i-1)}_{mu} (E)}.
\end{equation}
The first iteration is a canonical simulation at sufficiently
high temperature $T_0 = \frac{1}{\hat{\beta}_0}$,  
with 
$w^{(1)}_{mu}(E) = e^{-\hat{\beta}_{0}E}$.
For details, see Refs.~\cite{HO94_3} and \cite{HO95a}. The method for
calculating the multicanonical weight factor is by no means 
unique.  While it
is quite general, it has the disadvantage that it requires  
iterations of short simulations, the number of which is not known
{\it a priori}. 

Performing simulations in a so-called {\it $1/k$-ensemble}  is 
a  way of sampling the (microcanonical) {\it entropy} $S$ uniformly:
\begin{equation}
P_{1/k} (S) = {\rm const}~.
\lab{eq1k}
\end{equation}
This equation implies that 
\begin{equation}
P_{1/k} (E) = P_{1/k} (S) \frac{dS}{dE} \propto \frac{dS}{dE} 
= \tilde{\beta}(E)~,
\label{1overk}
\end{equation}
where $\tilde{\beta}(E)$ stands for the {\it effective}
inverse temperature\footnote{ 
We remark that the term ``multicanonical algorithm'' was inspired by
this effective temperature. In the early work, Berg 
and co-workers used for the
multicanonical weight the specific parametrization 
$\exp(-\tilde{\beta}(E) ~E- \alpha (E))$, approximating 
$S(E) = \log n(E)$ by straight lines between adjacent 
energy bins.}
and is defined by
\be
\tilde{\beta}(E) \equiv \frac{1}{\tilde{T}(E)} 
 = \frac{d \log n(E)}{d E}~.
\label{T_E}
\ee
There is again no explicit temperature dependence in simulations of the
$1/k$ ensemble; it can be said that simulations in both multicanonical
and $1/k$ ensembles include 
information of {\it all} temperatures. If the simulation is restricted to
a certain energy interval, Eq.~(\ref{T_E}) defines the corresponding
temperature range (assuming $n(E)$ is a monotonous function of $E$)
over which reliable results can be obtained.

In their original paper, Hesselbo and Stinchcombe \cite{HS} 
proposed that 
configurations are assigned a weight
\begin{equation}
w_{1/k} (E) = \frac{1}{k(E)}~,
\lab{eq9}
\end{equation}
where the function $k (E)$ is defined as the integral of
density of states with respect to energy $E$:
\begin{equation}
k(E) = \int_{-\infty}^{E} dE^{\prime} ~n(E^{\prime})~.
\label{1kw}
\end{equation}
This implies that
\be
P_{1/k} (E) \equiv n(E) w_{1/k}(E) = \frac{n(E)}{k(E)}
=\frac{d \log k(E)}{dE}~.
\lab{eq11}
\ee
Since the density of states $n(E)$ is a rapidly increasing 
function of energy, we have
$\log k(E) \approx \log n(E)$ for wide range of values of $E$.  
Hence, Eqs.~(\ref{1overk}) and
(\ref{eq11}) are equivalent, and a random walk in the entropy
space is realized.  Since the entropy $S(E)$ is a monotonically
increasing function of energy, a random walk in entropy implies
a random walk in energy space (with more weight towards low-energy
region; compare Eqs.~(\ref{eq2}) and (\ref{eq9})).
Hence, a simulation in $1/k$-ensemble can escape from any energy
barrier.
Hesselbo and Stinchcombe claim
that the $1/k$-sampling is superior to the multicanonical algorithm in
the sense that a fewer number of sampling is necessary for the
former.\cite{HS}

In numerical work, energy is discretized and integrals are replaced
by sums. Again, the weight $w_{1/k}(E)$ is not {\it a priori} known 
and its estimator has
to be calculated. It is obvious that the method described above 
for determination of the 
multicanonical weight factor is also suited for the calculation
of the weight in the $1/k$-ensemble. In addition, it follows from the
definitions of the weights, Eqs.~(\ref{eq2}), (\ref{eq9}), 
and (\ref{1kw}), that one can calculate $w_{1/k} (E)$ from
$w_{mu}(E)$, 
and vice versa. Thermodynamic quantities at any
temperature can be calculated by Eq.~(\ref{rewt}) with
the re-weighting techniques of Eq.~(\ref{erw}), in which
$P_{mu}(E)$ and $w_{mu}(E)$ are replaced by $P_{1/k}(E)$ and
$w_{1/k}(E)$, respectively.

The third approach is 
{\it simulated tempering}, which was first introduced under the name 
{\it method of expanded ensembles} by Luyabartsev 
et al.\ \cite{L}, but
became more popular under the former name proposed by 
Marinari and Parisi
in Ref.~\cite{MP}. In this ansatz temperature itself becomes a 
dynamical variable. Temperature and configuration are both updated with a
weight:
\begin{equation}
w_{ST} (T,E) = e^{-E/T - g(T)}~,
\end{equation}
where the function $g(T)$ is chosen so that the probability distribution
of temperature is given by
\begin{equation}
P_{ST}(T) = \int dE~ n(E)~ e^{-E/T - g(T)} = {\rm const}~.
\lab{eq12}
\end{equation}
Hence, in simulated tempering the {\it temperature} is sampled
uniformly, while
simulations in multicanonical and $1/k$ ensembles respectively sample
energy and entropy uniformly.  A random walk in temperature space
is realized,
allowing the simulation to escape from any energy barrier.
The last equation, Eq.~(\ref{eq12}), implies that
\begin{equation}
e^{g(T)} \propto \int dE~ n(E)~ e^{-E/T}~. \label{con}
\end{equation}
The function $g(T)$ is therefore proportional to the logarithm of the
canonical partition function at temperature $T$.  Again, the weight
$w_{ST}(T,E)$ is not {\it a priori}
known and its estimator has to be
calculated. We remark, however, that Eq.~(\ref{con}) allows easily the
calculation of the function $g(T)$ and therefore of the weight 
$w_{ST} (T,E)$, once the multicanonical weight $w_{mu} (E) = n^{-1} (E)$
is given. Likewise, the multicanonical weight $w_{mu} (E)$
can in principle be calculated from $g(T)$ by the inverse Laplace 
transformation. 

In the numerical work the temperature is discretized and 
restricted to a certain 
interval $[T_{min},T_{max}]$. Integrals are replaced by sums.
We also found it convenient to choose the temperature points $T_i$ not
equidistant, but so that the increment of adjacent temperature points
decreases 
exponentially with decreasing temperature. 
 Of course, this implies that
 we do not sample in a uniform way in temperature but by a monotone function of
 temperature.

Given the parameters $g_i = g(T_i)$ a simulated tempering simulation  may
use convential Metropolis algorithm \cite{Metro} with 
 two kinds of Monte Carlo updates:
\begin{itemize}
\item  Updates of configurations (conformations) at a fixed 
       temperature $T_i$ with probability \\ 
       $\min\{1,\exp [-(E_{new}-E_{old})/T_n ] \}$.
\item  Updates of temperatures with a fixed configuration (conformation)
       with probability \\
  $\min\{1,\exp[-E (1/T_{new} - 1/T_{old}) - (g(T_{new}) - g(T_{old})) ] \}$.
\end{itemize}
Using these updates, one can perform a simulated tempering simulation. 

Crucial for this method is again the determination of the 
parameters $g_i$ ($i=1, \cdots, n$).
Given the temperature points $T_i$ ($i=1, \cdots, n$ with
$T_1=T_{max} \ge T_i \ge T_n=T_{min}$), $g_i$
can be calculated by the following iterative procedure:
\begin{enumerate}
\item Start with a short canonical simulation of $m_1$ MC sweeps 
      updating only configurations at 
      temperature $T_1 = T_{max}$ and calculate the average
      energy $<E>_{T_1}$.  Formally, this can be regarded as a
      simulated tempering simulation at a fixed temperature $T_1$ with weight
      $w_{ST} (T,E) = e^{-E/T_1 -g_1}$ and $g_1=0$.
\item Calculate new parameters $g_j$ according to:
      \begin{equation}
      g_j =  \left\{ \begin{array}{ll}
             g_j + \log(m_j)~, &  1 \le j \le i \cr
             g_j +  
<E>_{T_i}~\left(\dis{\frac{1}{T_{i+1}} - \frac{1}{T_{i}}} \right)~, &
              j = i+1 \cr
             \infty~, & j > i +1
                      \end{array} \right.
      \end{equation}
\item Start a new simulation, now updating both configurations and
temperatures, with weight $w_{ST} (T,E) = e^{-E/T_j - g_j}$ 
      and sample the
      distribution of temperatures $T_j$ in the histogram $m_j= m(T_j)$. For
      $T = T_{i+1}$ calculate  the average energy $<E>_{T_{i+1}}$. 
\item If the histogram $m_j$ is approximately flat in the temperature range
      $T_1 \ge T_j \ge T_{i+1}$, set $i = i + 1$.  Otherwise, leave
      $i$ unchanged.
\item Iterate the last three steps until the obtained temperature 
      distribution $m_j$ becomes flat over the whole temperature range
      $[T_n=T_{min},T_1=T_{max}]$.
\end{enumerate}

Once the weight factor $w_{ST}(T,E)$ is obtained, we make a
production run with high statistics.  Physical quantities have to 
be sampled for each temperature point
separately. Their expectation values at temperature $T$ are then 
calculated in the usual way by
\begin{equation}
<{\cal O}>_T ~=~ \frac{\dis{\int dx~{\cal O}(x)~e^{-E(x)/T}}}
{\dis{\int dx~e^{-E(x)/T}}}~,
\lab{rewt2}
\end{equation}
where $x$ labels the conformations, and only those conformations
that were obtained at temperature $T$ are included in the integral.
Expectation values at intermediate temperatures can be calculated by
the reweighting techniques.\cite{FS}
 
\noindent
{\bf Peptide Preparation and Potential Energy Function}\\
Met-enkephalin has the amino-acid sequence Tyr-Gly-Gly-Phe-Met.
For our simulations the
backbone was terminated by a neutral NH$_2$-- ~group at the N-terminus
and a neutral~ --COOH group at the C-terminus as in the previous works of
Met-enkephalin.\cite{LS,SA3,RSA4,EnkO,HO}  The potential energy function
$E_{tot}$ that we used is given by the sum of
the electrostatic term $E_{es}$, 12-6 Lennard-Jones term $E_{vdW}$, and
hydrogen-bond term $E_{hb}$ for all pairs of atoms in the peptide together with
the torsion term $E_{tors}$ for all torsion angles:
\begin{eqnarray}
E_{tot} & = & E_{es} + E_{vdW} + E_{hb} + E_{tors},\\
E_{es}  & = & \sum_{(i,j)} \frac{332q_i q_j}{\epsilon r_{ij}},\\
E_{vdW} & = & \sum_{(i,j)} \left( \frac{A_{ij}}{r^{12}_{ij}}
                                - \frac{B_{ij}}{r^6_{ij}} \right),\\
E_{hb}  & = & \sum_{(i,j)} \left( \frac{C_{ij}}{r^{12}_{ij}}
                                - \frac{D_{ij}}{r^{10}_{ij}} \right),\\
E_{tors}& = & \sum_l U_l \left( 1 \pm \cos (n_l \chi_l ) \right),
\end{eqnarray}
where $r_{ij}$ is the distance between the atoms $i$ and $j$, 
and $\chi_l$ is the $l$-th torsion angle.
This $E_{tot}$ was used in the actual simulations of the three
algorithms.
The parameters ($q_i,A_{ij},B_{ij},C_{ij},
D_{ij},U_l$ and $n_l$) for the energy function were adopted 
from ECEPP/2.\cite{EC1}-\cite{EC3} 
The computer code
KONF90 \cite{KONF} was used. The peptide-bond
dihedral angles $\omega$ were fixed at the value 180$^\circ$
for simplicity,
which leaves 19 angles $\phi_i,~\psi_i$, and $\chi_i$ as 
independent variables. We remark that KONF90 uses a different convention 
for the implementation of the 
 ECEPP parameters (for example, $\phi_1$ of ECEPP/2 is equal to
 $\phi_1 - 180^{\circ}$ of KONF90).  Therefore our energy values are  
slightly different  
from those of the original implementation
of ECEPP/2.

\noindent
{\bf Computational Details}\\
Preliminary runs showed that all  methods  need roughly the same amount 
of CPU time for a fixed
number of MC sweeps (about 15 minutes for 10,000 sweeps
on an IBM RS6000 320H). Hence, we compared the different 
methods by performing
simulations with the same number of total MC sweeps. 
By setting this number to 1,000,000 sweeps we tried to ensure high 
 statistics. One MC sweep updates every torsion angle 
of the peptide once.

In the case of simulated tempering we chose 30 temperature points
which were exponentially distributed between $T_{min} = 50$ K and
$T_{max}=1000$ K, i.e.,~$T_i = T_{max} \times \gamma^{(i-1)}$ 
($i=1, \cdots, 30$) with 
$\gamma = (T_{min}/T_{max})^{1/29}$. This specific choice of 
temperature points was made because we found in preliminary runs that
for an equidistant distribution of temperatures we either needed a very 
large number of intermediate temperature points or
could not obtain reliable estimates for $g_i = g(T_i)$. Even with 
the above
choice of temperature points we needed 150,000 sweeps
to calculate the simulated tempering parameters $g_i$.

In our earlier work \cite{HO} we found that we needed 40,000 sweeps 
to calculate the
multicanonical weight for Met-enkephalin. To better compare the different
methods we tried to improve the weights by further iterations till the total
number of sweeps was again 150,000. Although the weight for 
$1/k$-ensemble can also be determined by the iterative procedure,
here we just 
calculated it from the obtained multicanonical weight by means of
Eq.~(\ref{1kw}) to save computation time.  

All thermodynamic quantities were then  calculated from
 a production run of 1,000,000 MC sweeps for each of the three methods 
which followed 10,000 sweeps
for thermalization. At the end of every
second sweep we stored the energy of the actual conformation 
 for future analysis of thermodynamic quantities. 
In all cases, the simulations started from completely random initial
conformations (``Hot Start'').

\noindent
{\bf RESULTS AND DISCUSSION} \\
We start the Results section by 
demonstrating some of the basic 
features of the three methods. 
As explained above, we expect for all three algorithms that
simulations result in a (weighted) random walk in energy space. 
In Fig.~1 we show the time series of energy from the production runs 
of 1,000,000 MC sweeps for the three methods.  They all exhibit
a random walk between low energy states and high energy states.
In Ref.~\cite{EnkO} it was shown that with the energy parameters
of KONF90, conformations with potential energies less than $-11$ kcal/mol
essentially have the same structure (with small deviations in
dihedral angles), which is 
the conformation with
the global-minimum energy.  The random walks in Fig.~1 all reached
this lowest-energy state many times.  
The numbers of independent
such visits will be examined in detail below.

By comparing Eqs.~(\ref{eq2}) and (\ref{eq9}), one expects that
the random walk in $1/k$-ensemble is supposed to be
biased towards the low-energy region, while that in multicanonical
ensemble does not have any bias (free random walk).
This tendency is apparent in the random walks in
Fig.~1; that of $1/k$-ensemble spends more time in the low-energy
region.  

In Fig.~2 histograms of energy distribution for the three methods
are shown.
As expected, the distribution is essentially flat for 
multicanonical ensemble, it increases with decreasing 
energy for $1/k$-ensemble.
As implied by Eq.~(\ref{1overk}), $P_{1/k}(E) \times \tilde{T}(E)$
is a flat curve (data not shown). Here, $\tilde{T} (E)$ is the
effective temperature defined in Eq.~(\ref{T_E}).
The distribution of energies for simulated tempering given in Fig.~2
is also flat, but has the tendency to increase as energy
decreases. The flat distribution does not imply uniform sampling of
temperature, since our temperature points are not equidistant. For
equidistant temperature points we would expect a distribution which
is proportional to the reciprocal of the specific heat.

In $1/k$-ensemble a free random walk in entropy space is expected
to be realized (see Eq.~(\ref{eq1k})).
In Fig.~3a the time series of the function $S(E) = \log n(E)$
is shown.  A random walk between small $S$ and large $S$ is indeed observed.
In Fig.~3b we display the probability distribution of entropy for
both simulations of multicanonical and $1/k$ ensembles.
This should be compared with Fig.~2,
where we displayed histograms of energy distribution. The
multicanonical simulation yields a curve
which increases with entropy, while in the $1/k$ ensemble we
observe a flat distribution, indicating that entropy was indeed
uniformly sampled.

In simulated tempering the
weight is chosen so that a 1d random walk in the temperature points $T_i$ is
obtained (see Eq.~(\ref{eq12})). This can be seen in Fig.~4, where 
we display the time series
of temperature for simulated tempering. As expected, a random walk
between high and low temperatures is observed.  Since all 
temperatures should
appear with the same weight, we expect that the distribution 
of temperatures 
 is essentially flat.  The simulation results confirmed this within the
deviations of factor 2 from flatness (data not shown).
As in the case of the
 multicanonical approach, one has to make a trade-off between 
the numerical
 effort one is willing to put into determination of weights and the
 deviation from a flat probability distribution one is willing to accept.
We allowed for differences of less than one order of magnitude. 

A major advantage of all three methods studied in this paper is 
that they allow calculations of thermodynamic quantities at any
temperature from just one simulation run, once
the weights are determined (see Eqs.~(\ref{rewt}) and
(\ref{rewt2})). 
As examples we show in Fig.~5 the average 
potential energy $<E>_T$ and the specific heat $C$ 
as functions of temperature.  The latter quantity is defined by
\begin{equation}
  C(T)  =  \frac{<E^2>_T - <E>_T^2}{N~T^2}~, 
\label{sph}
\end{equation}
where $N (=5)$ is the number of amino acids in the peptide.
All three methods yield  the same values at most of
the temperature values within the
(small) error bars. 
These results demonstrate that the three methods are 
equally well suited
for calculation of thermodynamic quantities. It  also proves 
that our estimates
of these quantities are reliable (since they were calculated 
from independent simulation data obtained by
different methods; there is no systematic hidden bias). This is especially 
important for the low temperature region where comparison with
canonical simulation is not possible.
Note that the average potential energy at the lowest-temperature
region is about $-12$ kcal/mol, which is the global-minimum 
energy value for the energy function in KONF90.\cite{EnkO}
The value at a high temperature, say $T=1000$ K, is as large
as $\approx 16$ kcal/mol.  The energy fluctuation $\delta E$
at this temperature
is about 5 kcal/mol (calculated from the value of $C$ from
Eq.~(\ref{sph}) by $\delta E = \sqrt{5 C/\bhat^2}$).  Hence, the 
random walks in Fig.~1 have
reached the high-energy region with energy values that would
be obtained at temperatures higher than 1000 K.

The behavior of the specific heat is also reasonable.  The peak
at $T \approx 300$ K implies that this temperature is 
most relevant for the folding of the peptide.  The zero-temperature
limit of the specific heat agrees with that of the harmonic 
approximation to the potential energy, where equipartition 
theorem can be used:
\be
<E> = N_F \frac{T}{2}~.
\lab{eqp}
\ee
Here, $N_F$ is the number of degrees of freedom (number of
torsion angles) and we have set
again $k_B = 1$.  The above specific heat is defined to be
the derivative of average energy per residue with respect to
temperature:
\be
C = \frac{d (<E>/N)}{dT} = \frac{N_F}{2~N}~.
\lab{eqpc}
\ee
For Met-enkephalin $N_F = 19$ (number of free torsion angles)
and $N=5$, and we have $C=1.9$.  The zero-temperature limit
extrapolated in Fig.~5 agrees with this value.

A comparison of Figs.~1 and 4 
gives us a way of comparing the performance of the three methods
in terms of sampling the ground-state conformations. It is evident
that low energy (temperature) states which are separated in the 
time series by
high energy (temperature) states 
are uncorrelated. The number of such ``tunneling'' events is therefore
a lower limit for the number of independent low-energy states visited
in the simulation. 
We define a tunneling event as the walk
in the simulation between a conformation with energy $E \le -11.0$ kcal/mol
(i.e.,~a ground-state conformation) (or $T=50$ K in the case
of simulated tempering) to a conformation with
energy above $E=20.0$ kcal/mol (or $T=1000$ K for simulated
tempering) and back to the ground-state region. 
For each tunneling event the lowest energy, $E_{GS}$, obtained
during the corresponding cycle was monitored.
Table 1 summarizes our results. 
In our production runs of each 1,000,000 MC sweeps we
observed 23 tunneling events for the multicanonical simulation,
27 for the simulation in the $1/k$-ensemble, and  19 in the case of
simulated tempering. These numbers correspond to tunneling times
(average number of MC sweeps needed for a tunneling event) 
of $40324$, $35664$, and $47874$ MC sweeps for multicanonical, 
$1/k$-ensemble, and simulated tempering simulations.  
Note that the
number of tunneling events from the multicanonical simulation is
higher than the one quoted in Ref.~\cite{HO94_3}, where we found only 
18 tunneling events and a tunneling time of 54136.
 We assume that this improvement is due to the
 improvement of the multicanonical weight.
It is interesting to observe that the number of tunneling events
(and therefore independent ground-state conformations found in the course
of simulation) is larger for the $1/k$ ensemble than for the multicanonical
simulation. 
The observed differences may be due to large statistical
fluctuations (see the standard deviations of tunneling times in the last row
of Table~1). 
In any case the
differences in tunneling times which we found are too small to establish a
ranking of the performances among the three methods. 

 We conclude that all three methods,
 while not differing much from each other, are much more efficient in finding 
 independent ground-state structures than traditional methods.
In Ref.~\cite{HO94_3} we made a comparison of multicanonical algorithms 
with simulated annealing, including annealing versions of the
multicanonical simulations.\cite{LC} Multicanonical annealing simplifies
the determination of weight factors and can be used to search groundstates, 
but does not allow estimation of thermodynamic quantities. 
Here we also performed annealing simulations in $1/k$-ensemble and
simulated tempering.  Twenty independent annealing runs with
50,000 MC sweeps were made for the two methods.  
An annealing simulation in $1/k$-ensemble
was done as in the same manner as for multicanonical annealing.
\cite{HO94_3}  Namely, 
an upper bound in energy $E_{wall}$ 
is introduced, above which a simulation is not allowed to enter.  The
weight is updated so that the energy distribution
becomes flat in the interval ($E_{wall} - \Delta E, E_{wall}$),
where $\Delta E$, sampling energy interval, is a constant.
The upper bound $E_{wall}$ is lowered once after each iteration so
that $E_{wall} = E_0 + \Delta E$, where $E_0$ is 
the lowest energy found in the preceeding iteration.
A simulated tempering annealing simulation is implemented similarly
by replacing energy by temperature in the above procedure
(for details of the annealing algorithms, see
Ref.~\cite{HO94_3}). 

 The lowest energies obtained by each annealing run
are summarized for the three methods  and compared with simulated annealing 
in Table~2, where the values for
multicanonical annealing and simulated annealing 
were taken from Ref.~\cite{HO94_3}.
The results are all similar with high probability (
75 -- 90 \% ) of finding the
energy global minimum in contrast with the moderate probability
(around 40 \%) of those by Monte Carlo simulated annealing
without much fine-tuning of the annealing schedule.\cite{HO94_3}
To compare different annealing schedules we varied the number of
independent runs and MC sweeps per run keeping their product constant.
The results are shown in Table~3, where data for
multicanonical annealing and simulated annealing 
were taken from Ref.~\cite{HO94_3}.
Again our results do not allow a ranking of the generalized ensemble
algorithms, but shows that they perform better than simulated annealing.

In order to obtain an optimal annealing schedule, one has 
to monitor the specific
heat.  One has to lower the temperature very slowly where the specific heat
has a peak so that a wide variety of conformations are sampled. 
>From Fig.~5 one finds that the present system of 
Met-enkephalin has a peak
of specific heat at $\approx 300$ K.  This temperature in turn
corresponds to the average potential energy of 
$\approx -1$ kcal/mol, as is shown in Fig.~5.  Since the
lowest energy is about $-12$ kcal/mol, we conclude that the
sampling energy interval $\Delta E$ \cite{HO94_3} for the annealing 
simulations in multicanonical and $1/k$ ensembles should be
more than 11 $(-1 - (-12))$ kcal/mol.  The results in Table~2 
were obtained
with $\Delta E = 15$ kcal/mol.  As for the simulated tempering
annealing simulation, we conclude that the corresponding 
sampling temperature interval $\Delta T$ should be
at least $\Delta T = 250$ $(300 - T_{min})$ K.  The results 
in Table~2 were
obtained with $\Delta T = 300$ K.  These analyses give some
hint for the determination of optimal annealing conditions.

Finally, in Fig.~6 we show the time series of energy from 
typical annealing runs
for multicanonical algorithm, $1/k$-ensemble, simulated tempering,
and simulated annealing.  From the Figure we can understand why
the performances of the former three algorithms were better than
a regular simulated annealing with naive exponential annealing
schedule of Ref.~\cite{KONF} (without reheating).  Namely,
as the simulation proceeds, the size of energy fluctuations is
fixed to be a preset finite value for the former three methods,
while it decreases towards zero for the latter.  Hence, the
chance of getting trapped in an energy local minima becomes
higher and higher as the simulation proceeds for Monte Carlo
simulated annealing.

\noindent
{\bf CONCLUSIONS} \\
We have performed simulations with high statistics for a simple
peptide, Met-enkephalin, to compare numerically the performances of three
new algorithms in the protein folding problem. All three methods have
in common that simulations are performed in an extended ensemble 
instead of the
ususal canonical ensemble. They all require as a first step 
the calculation
of estimators for the not {\it a priori} known probability 
weight factors. Using an analytical
transformation of the obtained distribution to a canonical distribution,
any thermodynamic quantity at any temperature 
can in principle be calculated from one
simulation run.
We demonstrated that
the calculated thermodynamic quantities over a wide range of
temperatures were identical for the three methods.
In particular, the agreement persisted into 
the difficult, low-temperature regime, where 
regular canonical simulations
will necessarily get trapped in one of huge number of energy
local minima.  Hence, one can cross-check the low-temperature results obatined
from one of the three methods by those from another.
Furthermore, by comparing the efficiency of the three methods in
finding independent ground-state structures,
we found that the three methods are equally more effective than
traditional methods.

We conclude that all three methods are equally well suited for the
simulation of peptides and proteins.
Temperature, energy, or entropy is also by no means
the only variables in which a uniform sampling is possible,
 nor is there
any restriction on one variable.
The
three studied algorithms are only special cases of a larger class of
similar methods, which we may call "simulations in 
{\it generalized ensemble}".

\vspace{0.5cm}
\noindent
{\bf Acknowledgements}: \\
Our simulations were 
performed on   a cluster of fast RISC workstations at SCRI (The Florida
State University, Tallahassee, USA) and computers at the Computer Center
of the Institute for Molecular Science (IMS), Okazaki,
Japan.
This work is
supported, in part, by the Department of Energy, contract DE-FC05-85ER2500,
by the 
Schweizerische Nationalfonds (Grant 20-40'838.94), and
by Grants-in-Aid for Scientific Research from the
Japanese Ministry of Education, Science, Sports, and Culture.
Part of this work was 
done while one of us (U.H.) was a visitor to IMS. U.H.\ would like to thank
the members of IMS for the kind hospitality extended to him and the  
Japan Society for the Promotion of Science (JSPS) for a 
generous travel grant.\\


 \newpage
 \noindent
 {\bf \Large TABLE CAPTIONS:}\\
{Table~1: Estimated ground-state energies $E_{GS}$ (in kcal/mol)
of each tunneling event
as obtained by the three algorithms. $t_{min}$ is the sweep 
 when the simulation first entered the groundstate region ($E \le
-11.0$ kcal/mol) in the corresponding tunneling event.
$<t_{tun}>$ is the average time in Monte Carlo sweeps
between these tunneling events. The numbers in brackets are the 
standard deviation of the corresponding quantities.}\\
{Table~2: Lowest energy (in kcal/mol) obtained by  the 20 annealing runs
of the three algorithms. For comparison, the results from simulated annealing
are also included.  For all cases, the total number of 
Monte Carlo sweeps 
per run was 50,000.
$n_{GS}$ is
the number of runs in which a conformation with $E \le -11.0$ kcal/mol
was
obtained. The data for multicanonical annealing and simulated annealing
were taken from 
Ref.~\cite{HO94_3}.}\\
{Table~3: Number of runs that reached ground-state conformations for
the three algorithms. Keeping the total number of MC sweeps constant
we varied the number of independent runs. For comparison, simulated
annealing data are included. The data for multicanonical annealing and 
simulated annealing
were taken from 
Ref.~\cite{HO94_3}.}\\

\newpage
Table~1.\\
\begin{table}[h]
\begin{center}
{\small
\begin{tabular}{||c||c|c||c|c||c|c||}\hline \hline
& \multicolumn{2}{c||}{Multicanonical Ensemble} &
  \multicolumn{2}{c||}{$1/k$-Ensemble} &
  \multicolumn{2}{c||}{Simulated Tempering} \\ \hline
$n_{tu}$ & $t_{min}$ & $E_{GS}$ & $t_{min}$ & $E_{GS}$ & $t_{min}$ & $E_{GS}$  
 \\ \hline
 0 &  24830& $-11.8$ &    2162& $-11.8$&    12596& $-12.0$\\
 1 &  49472& $-11.7$&   29094& $-11.7$&   66286& $-11.8$\\
 2 & 160546& $-12.0$&   56382& $-12.0$&  106646& $-11.9$\\
 3 & 228340& $-11.9$&   85790& $-12.0$&  163430& $-12.0$\\
 4 & 240648& $-11.9$&  121798& $-12.0$&  183162& $-11.9$\\
 5 & 272908& $-11.8$&  168464& $-11.6$&   237592& $-11.7$\\
 6 & 294894& $-11.3$&  174704& $-11.8$&   273200& $-11.9$\\
 7 &  319722& $-12.1$&  209170& $-12.0$&  408646& $-11.3$\\
 8 &  330868& $-11.1$&  236218& $-11.9$&  446152& $-11.8$\\
 9 &  484056& $-11.7$&  299008& $-12.0$&  455142& $-11.8$\\
10 &  503570& $-11.9$&   341488& $-12.1$&  465150& $-11.9$\\
11 &  549780& $-11.7$&  389804& $-12.0$&  514524& $-11.8$\\
12 & 623528& $-11.5$&  450304& $-12.1$&  639764& $-11.3$\\
13 &  628398& $-11.8$&  500868&$-11.7$&  648672& $-11.4$\\
14 & 651510& $-11.1$&  506178& $-11.9$&  665176& $-11.9$\\
15 &  663014& $-11.5$&  516248& $-11.9$&  691068& $-11.8$\\
16 &  681694& $-11.9$&  578230& $-12.0$&   746726& $-11.9$\\
17 &  694916& $-11.3$&  588002& $-11.6$&   778272& $-12.0$\\
18 &  815442& $-11.8$&  638984& $-12.0$&  899104& $-11.8$\\
19 &  885534& $-11.1$&  691742& $-11.2$& 922208& $-11.8$\\
20 &  911926& $-11.6$&  696882& $-11.9$& &\\
21 &  918322& $-11.7$&  705466& $-11.8$& &\\
22 &  949836& $-11.5$&  759824& $-11.9$& &\\
23 &  952284& $-11.2$&   803662& $-12.1$& &\\
24 &        &       &  875652& $-11.6$& &\\
25 &        &       &  890076& $-12.0$& &\\
26 &        &       &922426& $-11.9$& &\\
27 &        &       & 965106& $-12.0$& &\\ \hline
$<t_{tun}>$& \multicolumn{2}{c||}{40324(39826)} &
             \multicolumn{2}{c||}{35664(19633)} &
             \multicolumn{2}{c||}{47874(37777)}\\
\hline \hline
\end{tabular}}
\end{center}
\end{table}

\newpage
Table~2.\\
\begin{table}[h]
\begin{center}
{\small
\begin{tabular}{||c||c|c|c|c||}\hline \hline
Run& Multicanonical & ~~~~$1/k$~~~~ & Simulated Tempering & Simulated 
Annealing \\ \hline
1  & $-11.6$ & $-10.3$ & $-11.8$ &$-11.7$\\
2  & $-12.0$ & $-11.7$ &$-11.7$ &$-8.6$\\
3  & $-10.2$ & $-11.3$ &$-11.5$ &$-12.1$\\
4  & $-10.1$ & $-11.6$ &$-11.4$ &$-8.8$\\
5  & $-11.9$ & $-12.1$ &$-11.1$ &$-7.4$\\
6  & $-12.0$ & $-11.8$ &$-11.1$ &$-8.9$\\
7  & $-11.9$ & $-10.9$ &$-11.6$ &$-12.1$\\
8  & $-11.7$ & $-10.4$ &$-11.5$ &$-12.2$\\
9  & $-11.8$ & $-11.7$ &$-11.4$ &$-7.1$\\
10 & $-11.9$ & $-11.4$ &$-11.5$ &$-7.5$\\
11 & $-12.0$ & $-11.3$ &$-11.3$ &$-9.9$\\
12 & $-12.1$ & $-11.9$ &$-10.5$ &$-7.3$\\
13 & $-12.0$ & $-11.4$ &$-12.0$ &$-8.4$\\
14 & $-11.8$ & $-12.0$ &$-9.0$ &$-10.6$\\
15 & $-11.3$ & $-11.6$ &$-11.8$ &$-10.3$\\
16 & $-11.9$ & $-11.5$ &$-10.9$ &$-12.2$\\
17 & $-12.0$ & $-11.7$ &$-11.4$ &$-12.2$\\
18 & $-11.9$ & $-9.3$ &$-10.7$ &$-9.1$\\
19 & $-11.9$ & $-11.6$ &$-11.1$ &$-11.9$\\
20 & $-11.6$ & $-10.5$ &$-11.7$ &$-12.1$\\ \hline
$n_{GS}$& 18/20 & 15/20& 16/20& 8/20
\\ \hline \hline
\end{tabular}
}
\end{center}
\label{tab2b}
\end{table}

\newpage
Table~3.\\
\begin{table}[h]
\begin{center}
{\small
\begin{tabular}{||c||c|c|c|c||}\hline \hline
Number of Runs& Multicanonical & ~~~~$1/k$~~~~ & Simulated Tempering & 
Simulated Annealing \\ 
$\times$ Number of MC Sweeps& & & & \\ \hline 
$10 \times 100,000$ & 10 & 9 & 9 & 5\\
$20 \times 50,000$  & 18 & 15 & 16 & 8\\
$50 \times 20,000$  & 21 & 22 & 17 & 10\\
$100 \times 10,000$ & 28 & 29 & 20 & 13
\\ \hline \hline
\end{tabular}
}
\end{center}
\label{tab3}
\end{table}

\newpage
\noindent
{\bf \Large FIGURE CAPTIONS:}\\
FIG.~1: Time series of energy $E$ (kcal/mol) 
from a simulation of 1,000,000 MC sweeps by multicanonical,
$1/k$-sampling, and simulated tempering algorithms.\\
FIG.~2: Probability distribution of energy $E$ (kcal/mol)
from simulations by the three methods.
	 The data rely on 1,000,000 MC sweeps
	for each simulation.\\
FIG.~3a: Time series of entropy $S$ from a $1/k$-ensemble simulation of
	1,000,000 MC sweeps.\\
FIG.~3b: Probability distribution of entropy $S$ from 
simulations in multicanonical
	and $1/k$ ensembles. The data rely on 1,000,000 MC sweeps
	for each simulation.\\
FIG.~4: Time series of temperature $T$ (K)
from a simulated tempering simulation of
	1,000,000 MC sweeps.\\
FIG.~5: Average energy $<E>_T$ (kcal/mol)
and specific heat $C$ 
as a function of temperature $T$ (K) calculated
        from the simulations
 	 of multicanonical ensemble, $1/k$-ensemble, and simulated tempering. 
         The data rely on 1,000,000 MC sweeps for each method.
The energy scale is displayed on the ordinate on the left-hand side
and the specific heat on the right-hand side.  The average energy is
a monotonically increasing function of temperature, whereas the
specific heat has a maximum around 300 K.\\
FIG.~6: Time series of energy $E$ (kcal/mol)
         from an annealing simulation 
         of 50,000 MC sweeps by multicanonical, $1/k$-sampling, 
simulated tempering, and simulated annealing algorithms.\\


\noindent
\begin{thebibliography}{(00)}
\bibitem{VNS} M.~V{\'a}squez,\ G.~N{\'e}methy and 
H.A.~Scheraga, {\it Chem.~Rev.}, {\bf 94}, 2183 (1994).
\bibitem{LS} Z. Li and H.A. Scheraga, 
{\it Proc.~Nat.~Acad. Sci.~U.S.A.},
{\bf 84}, 6611 (1987).
\bibitem{Fo} For a review see~ S.~Forrest, {\it Science}, {\bf 261},
             872 (1993).
\bibitem{SA} S. Kirkpatrick, C.D. Gelatt, Jr. and M.P. Vecchi,  
  {\it Science}, {\bf 220}, 671 (1983).
\bibitem{SA1} M. Nilges, A.M. Gronenborn, A.T. Br{\"u}nger,
and G.M. Clore, {\it Protein Engng}, {\bf 2}, 27 (1988).
\bibitem{SA2} S.R. Wilson, W. Cui, J.W. Moskowitz, and K.E. Schmidt,
{\it Tetrahedron Lett.}, {\bf 29}, 4373 (1988).
\bibitem{SA3} H. Kawai, T. Kikuchi, and Y. Okamoto, {\it Protein
Eng.}, {\bf 3}, 85~(1989).
\bibitem{SA4} C. Wilson and S. Doniach, {\it Proteins}, {\bf 6},
193 (1989).
\bibitem{SA5} S.R. Wilson and W. Cui, in {\it The Protein Folding
Problem and Tertiary Structure Prediction}, K.M. Merz, Jr. and
S.M. Legrand, eds. (Birkh{\"a}user, 1994) pp. 43--70.
\bibitem{MU} B.A. Berg and T. Neuhaus, {\it Phys. Lett.},
  {\bf B267}, 249 (1991); {\it Phys. Rev. Lett.},
  {\bf 68}, 9 (1992).
\bibitem{MU3} B.A. Berg, {\it Int.~J.~Mod.~Phys.}, {\bf C3},
  1083 (1992).
\bibitem{ES} J. Lee, {\it Phys. Rev. Lett.}, {\bf 71}, 211 (1993);
{\it Phys. Rev. Lett.}, {\bf 71}, 2353(E) (1993).
\bibitem{L}  A.P.~Lyubartsev,~A.A.Martinovski,\ S.V.~Shevkunov, and \
	     P.N.\ Vorontsov-Velyaminov,\ {\it J.~Chem.~Phys.}, {\bf 96},
	     1776 (1992).
\bibitem{MP} E.~Marinari and G.~Parisi, {\it Europhys.~Lett.}, {\bf 19},
	     451 (1992).
\bibitem{HS} B.~Hesselbo and R.B.~Stinchcombe,\ {\it Phys.~Rev.~Lett.},
	     {\bf 74}, 2151 (1995).
\bibitem{COMM} B.A. Berg, U.H.E. Hansmann, and Y. Okamoto,  
  {\it J. Phys. Chem.}, {\bf 99}, 2236 (1995); but see also
the reply by M.H. Hao and H.A.  Scheraga, {\it J. Phys. Chem.},
{\bf 99}, 2238 (1995).
\bibitem{HO} U.H.E. Hansmann and Y. Okamoto, {\it J.~Comp.~Chem.}, 
  {\bf 14}, 1333 (1993).
\bibitem{HO94_3} U.H.E. Hansmann and Y. Okamoto, {\it J. Phys. Soc. Jpn.},
{\bf 63}, 3945 (1994); {\it Physica A}, {\bf 212}, 415 (1994).
\bibitem{HSp} M.H. Hao and H.A.  Scheraga, {\it J. Phys. Chem.},
 {\bf 98}, 4940 (1994).
\bibitem{HO95a} Y.~Okamoto, U.H.E.~Hansmann, and T. Nakazawa,\ 
{\it Chem. Lett.}, {\bf 1995}, 391;
Y.~Okamoto and U.H.E.~Hansmann,\ {\it J.~Phys.~Chem.},
		{\bf 99}, 11276 (1995).
\bibitem{IP} A. Irb{\" a}ck and F. Potthast, {\it J. Chem. Phys.}, 
{\bf 103}, 10298 (1995).
\bibitem{CA} R.H. Swendsen and J.S. Wang, {\it Phys. Rev. Lett.},
  {\bf 58}, 86 (1987).
\bibitem{CA2} R.G. Edwards and A. Sokal, {\it Phys. Rev.},
  {\bf D38}, 2009 (1988).
\bibitem{FS} A.M. Ferrenberg and R.H. Swendsen, {\it Phys.\ Rev.\ Lett.},
  {\bf  61}, 2635 (1988); {\it Phys. Rev. Lett.}, {\bf 63 }, 
1658(E) (1989), and
  references given in the erratum.
\bibitem{Metro} N. Metropolis, A.W. Rosenbluth, M.N. Rosenbluth, 
  A.H. Teller, and E. Teller, {\it J. Chem. Phys.}, {\bf 21}, 
  1087 (1953).
\bibitem{RSA4} B. von Freyberg and W. Braun, {\it J. Comp. Chem.},
{\bf 12}, 1065 (1991).
\bibitem{EnkO} Y.~Okamoto, T.~Kikuchi, and H.~Kawai, {\it Chem. Lett.},
{\bf 1992}, 1275.
\bibitem{EC1} F.A. Momany, R.F. McGuire, A.W. Burgess, and H.A.
Scheraga, {\it J. Phys. Chem.}, {\bf 79}, 2361~(1975).
\bibitem{EC2} G. N{\'e}methy,
M.S. Pottle, and H.A. Scheraga, {\it J. Phys. Chem.}, {\bf 87},
1883~(1983).
\bibitem{EC3} M.J. Sipple, G. N{\'e}methy, and H.A. Scheraga,
{\it J. Phys. Chem.}, {\bf 88}, 6231~(1984).
\bibitem{KONF} H. Kawai, Y. Okamoto, M. Fukugita, T. Nakazawa, and  
  T. Kikuchi, {\it Chem. Lett.}, {\bf 1991}, 213;
  Y. Okamoto, M. Fukugita, T. Nakazawa, and 
  H. Kawai, {\it Protein Engng.}, {\bf 4}, 639 (1991).
\bibitem{LC} J. Lee and M.Y. Choi, {\it Phys.~Rev.}, {\bf E50},
  R651 (1994).

\end{thebibliography}
\end{document}